\documentclass[aps,preprint,amsmath,amssymb,showpacs]{revtex4}
\usepackage{graphicx, epsfig,color}

\def\be{\begin{eqnarray}}
\def\ed{\end{eqnarray}}
\def\bT{{\bf T}}
\def\ga{\gamma}
\def\non{\nonumber}
\def\ra{\rangle}
\def\la{\langle}

\begin{document}

\title{\Large \bf Lepton flavor violating tau decays in type-III seesaw mechanism}

\date{\today}

\author{ \bf  Abdesslam Arhrib$^{1}$\footnote{Email:aarhrib@ictp.it}
 }

\author{ \bf  Rachid Benbrik$^{2,3}$\footnote{Email:rbenbrik@mail.ncku.edu.tw}
 }

\author{ \bf  Chuan-Hung Chen$^{2,3}$\footnote{Email:physchen@mail.ncku.edu.tw}
}

\affiliation{$^1$ D\'epatement de Math\'ematique, Facult\'e des
Sciences and Techniques, Universit\'e Abdelmalek Essa\^adi, B. 416, Tangier, Morocco \\
%
$^{2}$
Department of Physics, National Cheng-Kung
University, Tainan 701, Taiwan \\
$^{3}$National Center for Theoretical Sciences, Hsinchu 300,
Taiwan
 }

\begin{abstract}

In this paper, the lepton flavor violating
$\tau\to \ell P(V)$ ($P,V= \pi^0, \eta, \eta^{\prime}, \rho^0, \omega, \phi$) and $\tau\to 3\ell$ ($\ell = e, \mu$) decays are studied
in the framework of the type-III seesaw model, 
 in which new triplet fermions with a zero hypercharge ($Y=0$) interact
with ordinary lepton doublets via Yukawa couplings, 
and affect tree-level leptonic Z-boson couplings. 
We investigate the experimental bound from the leptonic Z
decay to get contraint on the exsiting  parameters space.
We predict that the upper limits on the branching ratios of
$\tau\to \ell P(V)$ and $\tau\to 3\ell$
can reach the experimental current limits.

\end{abstract}
\maketitle


\section{Introduction}

In the Standard Model (SM) with massless neutrinos, lepton flavor is
conserved. However, the current neutrino oscillation data
experiments indicate with very convincing evidence that neutrinos
are massive and lepton flavor are mixed \cite{Fukuda:1998mi}. This
is a powerful incentive for considering new particles
and interactions those of the Standard Model (SM) of quarks and
leptons. If the neutrino oscillation phenomenon takes place
actually, lepton flavor symmetry would be broken. In that case,
however, lepton flavor violating (LFV) processes are still highly
suppressed because of the smallness of neutrinos masses. Hence, any
experimental signal of charged LFV would be a clear indication of
physics beyond the SM. This fact has led to a great amount of
theoretical effect for revealing the underlying new physics in the
leptonic flavor sector.

LFV appears in various extensions of the SM. In particular LFV
decays $\tau \to 3\ell$ (where $\ell = e$ or $\mu$) are discussed in
various models
\cite{Ellis:2002fe,Saha:2002kt,Brignole:2004ah,Barbier:2004ez}. Some
of these models with certain combinations of parameters predict that
the branching fractions for $\tau \to 3\ell$ can be as high as
$10^{-7}$, which is already accessible in high-statistics B factory
experiments. Searches for LFV in charged lepton sector such as
$\tau\to \mu P (V)$ decays with a pseudoscalar or vector meson are
also discussed in models with Higgs mediated LFV processes
\cite{Sher:2002ew,Chen:2006hp}, 
heavy singlet Dirac
neutrinos \cite{Ilakovac:1999md}, dimension-six
effective fermionic operators that induce $\tau-\mu$ mixing
\cite{Black:2002wh}, R-parity violation in SUSY \cite{Li:2005rr},
type III two-Higgs doublet models \cite{Li:2005rr} and flavor
changing $Z^{\prime}$ bosons \cite{Li:2005rr}.
At the LHC, the $\tau$ leptons are
produced predominantly from decays of $B$ and $D$ mesons and $W$ and
$Z$ bosons. In the low-luminosity phase, corresponding to an
integrated luminosity of 10 fb$^{-1}$ per year, one expects
approximately $10^{12}$ and $10^{8}$ $\tau$ leptons produced per
year from heavy meson and weak boson decays, respectively
\cite{Unel:2005fj}. If we restrict the $\tau$ from weak bosons
decays only, and assuming that a branching ratio close to the
current upper limit, we can expect approximately 10 events within
the acceptance range of a typical LHC general purpose detector after
one year of low-luminosity running. With 30 fb$^{-1}$ of data, it
should be possible to probe branching ratios down to a level of
${\rm Br (\tau \to 3\ell, \ell P, \ell V)}$ $\approx$ $10^{-8}$ at
the LHC.

In order to give mass to the neutrinos, several ways have been
studied in \cite{type123,Mohapatra:1979ia,Foot:1988aq,Magg:1980ut}. An
alternative, equally valid and rather economical possibility is to
extend the lepton sector of the SM by a heavy triplet fermions and allow them
to interact with the ordinary lepton doublets via Yukawa couplings. In this scenario, the Higgs sector is unmodified, and a set of self-conjugate $SU(2)_L$ triplets of exotic leptons with zero hypercharge are added
that model so-called type III seesaw mechanism.

The model has many interesting features, including the
possibility of having low seesaw scale of order a TeV to realize leptogenesis \cite{Hambye:2003rt} and detectable effects at LHC\cite{Bajc:2007zf,Franceschini:2008pz} due to the fact that the heavy triplet leptons have
gauge interactions being non-trivial under the $SU(2)_L$ gauge group. In
particular, if kinematically accessible, the charged component of
the triplet will be produced in high energy collision, and its decay
into Higgs and light lepton\cite{Bajc:2007zf} provides a rather
spectacular signature.

Fermionic triplet effects have been studied in the lepton sector
\cite{Abada:2008ea} such as $\tau \to 3 \ell$, $\ell \to \ell^{\prime} \gamma$, $Z\to \ell \ell^{\prime}$, $\mu - e$ conversion and the
anomalous magnetic moment of leptons $(g-2)$\cite{Biggio:2008in, Kamenik:2009cb}. 
Several other processes have not been studied in the context of type
III seesaw model such as $\tau\to \ell P$ and $\tau\to \ell V$.

In this paper 
we try to demonstrate that we can have a contribution to lepton
flavour violating decays even with one triplet and singlet fermions. 
The paper is organized as follows: 
In the next section, 
we will recall the basic features of type-III seesaw model and discuss the motivations which adds a triplet. 
In Sec.~\ref{Constraints} 
we will discuss the constraint on the Yukawa couplings coming from neutrino experiments.
Section \ref{decay rates} then discusses the
analytical LFV $\tau$ decay rates with estimates the corresponding
LFV observables and conclusions are drawn in section \ref{conclusions}.
%
\section {Z-mediated LFV in the type-III seesaw model}
%
To study the lepton flavor violating effects in the so-called
type-III seesaw models
\cite{Mohapatra:1979ia,Foot:1988aq,Ma:2002pf}, we consider the
$SU(2)_L$ fermionic triplet with the quantum number of $(1,3,0)$ under $SU(3)_c\times
SU(2)_L\times U(1)_Y$ gauge symmetry \cite{Foot:1988aq}. 
For explaining the current data in neutrino physics, 
model with only one triplet fermion is not sufficient; 
therefore, more
fermionic triplets and/or singlets should be considered. 
Since our purpose is to illustrate the
$\tau$ LFV in the framework of type-III seesaw models, for simplicity we focus on the minimal extension of the SM (MSM), i.e. the case with one triplet and one singlet fermions \cite{Kamenik:2009cb}. The detailed analysis in the model with three triplets could be referred to Ref.~\cite{He:2009tf}.
%
Let us describe the model in more detail 
to identify new tree-level FCNC
in the lepton sector, the component fields  
of the triplet fermion is chosen to be
 \be
 \bT= \left(
           \begin{array}{cc}
              T^0/\sqrt{2}  & T^-\\
             T^+  & T^0/\sqrt{2} \\
           \end{array}
         \right)\,.
 \ed
In order to keep invariance under $SU(2)_L$ gauge transformation, we have
required the transformation of $\bT$ to be $\bT \to U^* \bT
U^\dagger$.
For studying flavor changing effects, we need to understand the
gauge couplings of SM leptons and triplet fermion. 
{The Yukawa sector with respect to the gauge symmetry
$SU(2)_L\times U(1)_Y$ 
can be written as \cite{Kamenik:2009cb}
 \be
-{\cal L}_Y &=& H^{\dagger} \bar e_{a}  Y^{ab}_E  L_b  +
y^a_T Tr \left( \bar T^c_L L_a H^T\right)+ y^a_S H^T i\tau_2 L_a S \non\\
&& + \frac{1}{2} m_T Tr
(\bar \bT \bT) \label{eq:Yuka} + \frac{1}{2} m_S S^T C S+h.c. 
 \ed
where $H^T=(\phi^+, \phi^0)$ is the SM Higgs doublet, $e_a$ denotes
right-handed lepton and $a (b)$ is the corresponding lepton flavor,
$L^T=(\nu, \ell)_L$ is the weak gauge doublet of lepton, $Y^{ab}_E$
and $y^a_{T,S}$ are Yukawa couplings, $C$ is the charge conjugation operator, and $m_{T(S)}$ is the mass of the new
stuff in triplet (singlet) fermion.}\\
Similarly, the relevant
gauge kinetic terms are written as
 \be
{\cal L}_{\rm kin}=\bar L i\not D_2 L + \bar \ell_R i\not D_1 \ell_R
-Tr[ \bar \bT'_L i\not D_3 \bT'_L], \label{eq:kin}
 \ed
where $D_{2\mu} =\partial_{\mu} + ig/2 \vec\tau \cdot \vec W_\mu
-ig'/2 B_\mu$, $D_{1\mu} =\partial_{\mu} -ig' B_\mu$ and $D_{3\mu}
=\partial_{\mu} + ig\vec\tau \cdot \vec W_\mu $ are the covariant
derivatives, $\bT'=i\tau_2 \bT$ and the associated gauge
transformation is $\bT'\to U \bT' U^\dagger$. 
Since singlet fermion doesn't couple to gauge boson, we don't show it in Eq.~(\ref{eq:kin}). Although charged
currents will induce flavor changing effects through box and penguin
diagrams, however, due to loop suppression, the Z-mediated LFV
induced at tree level will be dominant. Consequently, we focus on
the Z-boson related interactions. By Eq.~(\ref{eq:kin}), the
interactions in weak eigenstates of lepton are found by
 \be
  {\cal L}_{Z} = \frac{g}{2\cos\theta_W} \bar \ell
\ga^\mu \left( \cos2\theta_W X_L P_L - 2\sin^2\theta_W X_R P_R
\right) \ell Z_\mu \label{eq:Z}
 \ed
with $\ell_{L}=(e_L,\mu_L,\tau_L, T^c_R)$,
$\ell_{R}=(e_R,\mu_R,\tau_R, T^c_L)$ and
 \be
 X_L&=&\left(
             \begin{array}{cc}
            { \openone}_{3\times 3} & 0_{3\times 1} \\
             0_{1\times 3} & -2\cos^2\theta_W/\cos2\theta_W
             \end{array}
           \right)\,, \ \ \
 X_R=\left(
             \begin{array}{cc}
            { \openone}_{3\times 3} & 0_{3\times 1} \\
             0_{1\times 3} & \cos^2\theta_W/\sin^2\theta_W
             \end{array}
           \right)\,. \label{eq:XLR}
 \ed
Here, we have taken $T^c$ as the charge-conjugation state of $T^+$.
Since the couplings of the new charged leptons to Z-boson differ
from those of ordinary leptons, we will see that LFV at tree level
will be induced by the misalignment between weak and physical
states. To understand the effects of LFV, we first introduce two
unitary matrices $V_{L, R}$ that transform the weak states to
physical states by $\ell_{L(R)} \to V_{L(R)} \ell_{L(R)}$. Then the
matrices in Eq.~(\ref{eq:XLR}) become
 \be
 Z_\eta \equiv V_\eta X_\eta V^\dagger_\eta  = V_\eta I
V^\dagger_\eta + V_\eta (X_\eta-I) V^\dagger_\eta
 \ed
with $\eta=L, R$ and $I$ being the unit matrix. Immediately, we can
see that the lepton flavor changing effects are associated with
 \be
Z_{\eta ij} &=& \chi_\eta V_{\eta i4} V^*_{\eta j4 }\,, \non\\
\chi_L &=& -2\cos^2\theta_W/\cos2\theta_W -1 \,, \non\\
\chi_R &=& \cos^2\theta_W/\sin^2\theta_W-1. \label{eq:Zij}
 \ed
To get the information on $V_{\eta i4}$ and $V_{\eta 4i}$, we need
to study the detailed mass matrix for charged leptons.

After SSB, the mass matrix for charged lepton could be written as
 \be
 -{\cal L}^\ell_{Y}= \bar \ell_R M_{\ell}  \ell_L + h.c.
 \ed
with
 \be
 M_{\ell}= \left(
             \begin{array}{ccc}
          ({\bf Y}_{E})_{3\times 3}v/\sqrt{2}  & | & \bf 0_{3\times 1}
             \\
             --- & | & --\\
            ({\bf y^{\dagger}}_{T})_{ 1\times 3}v/\sqrt{2}   & | & m_T \\
             \end{array}
           \right)\,. \label{eq:Mll}
 \ed
Moreover, by choosing a suitable basis, indeed Eq.~(\ref{eq:Mll})
can be further simplified. For instance, using the transformation
$\ell_{L(R)}\to U_{L(R)} \ell_{L(R)}$ with
 \be
U_{L(R)}= \left(
             \begin{array}{ccc}
           \bar U_{L(R) 3\times 3} & | & \bf 0_{3\times 1}
             \\
             --- & | & --\\
             \bf 0_{1\times 3} & | & 1 \\
             \end{array}
           \right)\,,\non
 \ed
the matrix ${\bf Y}_{E}$ in Eq.~(\ref{eq:Mll}) can be diagonalized
and Eq.~(\ref{eq:Mll}) becomes
 \be
M_{\ell}= \left(
             \begin{array}{ccc}
          ({\bf m}_E)_{3\times 3}  & | & \bf 0_{3\times 1}
             \\
             --- & | & --\\
            ({\bf h}^\dagger_{T})_{ 1\times 3}  & | & m_T \\
             \end{array}
           \right)\,, \label{eq:mdia}
 \ed
where ${\rm diag}({\bf m}_{E}) = {\rm diag}(\bar U_{R} ({\bf
Y}_{E}v/\sqrt{2}) \bar U^\dagger_L)=(m_e, m_\mu, m_\tau )$ and ${\bf
h}_T= \bar U_{L}{\bf y}_T  v/\sqrt{2}$. Since $M_{\ell}$ still has
off-diagonal elements, clearly $m_{e,\mu,\tau}$ are not physical
eigenstates. In addition, from Eq.~(\ref{eq:mdia}) one can expect
that the lepton flavor violating effects will be associated with
${\bf h}_{T}$.
To get the physical states, we use $V_{R,\, L}$ introduced early to
diagonalize the mass matrix of lepton, i.e. $M^{\rm dia}_{\ell} =
V_{R} M_{\ell} V^\dagger_L$. The individual information on $V_L$ and
$V_R$ can be obtained by
 \be
 M^{\rm dia^\dagger}_{\ell} M^{\rm dia}_{\ell}  &=& V_{L}
 M^\dagger_{\ell} M_{\ell} V^\dagger_L \,,\non \\
M^{\rm dia}_{\ell}M^{\rm dia^\dagger}_{\ell}   &=& V_{R}
 M_{\ell} M^\dagger_{\ell} V^\dagger_R \label{eq:M_2}
 \ed
with
 \be
 M^\dagger_{\ell} M_{\ell} = \left(
             \begin{array}{ccc}
          {\bf m}^\dagger_E {\bf m}_E +{\bf h}_T {\bf h}^\dagger_T & | & {\bf h}_T m_T
             \\
             --- & | & --\\
             m_T {\bf h}^\dagger_{T}  & | & m^2_T \\
             \end{array}
           \right)\,, \ \ \
 M_{\ell} M^\dagger_{\ell} = \left(
             \begin{array}{ccc}
          {\bf m}_E {\bf m}^\dagger_E  & | & {\bf m}_E {\bf h}_{T}
             \\
             --- & | & --\\
              {\bf h}^\dagger_{T} {\bf m}^\dagger_E & | & m^2_T +{\bf h}^\dagger_{T}{\bf h}_{T}  \\
             \end{array}
           \right)\,.
           \non
 \ed
Clearly, $V_L$ and $V_R$ are the unitary matrices to diagonalize the
matrix $M^\dagger_\ell M_{\ell}$ and $M_{\ell} M^\dagger_{\ell}$,
respectively. Expectably, the off-diagonal elements of flavor mixing
matrices will be associated with $({\bf m}_{E ii}{\bf h}_{Ti})$ and
$m_T {\bf h}_{Ti}$ which reflect the mixture of ordinary quarks and
triplet fermion. Although in general the $4\times 4$ matrices will
be complicated and unknown, however, since the introduced triplet
fermions are much heavier than SM leptons, i.e. $m_T \gg {\bf m}_{E
ii}, {\bf h}_{Ti}\sim v$, for a good approximation we can expand
$V_{L(R)}$ to be $V_{L (R)} \approx { \openone}_{4\times 4 } +
\Delta_{L(R)}$ where $\Delta_{L(R)}$ is regarded as
$O(h_{Ti}/m_T)[O( m_{Eii} h_{Ti}/m^2_T)]$. Comparing with
Eq.~(\ref{eq:Zij}), we see $V_{\eta i4(4i)}\approx \Delta_{\eta
i4(4i)}$. From Eq.~(\ref{eq:M_2}), we can derive the leading order
for flavor mixing as
 \be
\Delta_{Li4}& \approx & \Delta^*_{L4i} \approx-\frac{m_T
h_{Ti}}{m^2_T - m^2_{Ei}-|h_{Ti}|^2 }\,,\\
\Delta_{Ri4}&\approx& -\Delta^*_{R4i}\approx -\frac{m_{Ei}
h_{Ti}}{m^2_T +{\bf h}^\dagger_{T}{\bf h}_{T} - m^2_{Ei}} \,.
 \ed
Since $m^2_T\gg m_T h_{Ti} \gg m_{Ei} h_{Ti}$, it is clear that the
effects of $\Delta_{R i4(4i)}$ are negligible. Hence, the
significant LFV in type-III seesaw model is only associated with
left-handed neutral currents.
\section{Constraints on the physical parameters}
\label{Constraints}
{In this section, we discuss the constraints coming from the 
neutrino experiments on the relevant Yukawa couplings $y^a_{T}$.  
%
%
%
After spontaneous symmetry breaking (SSB), where the Higgs field is
driven to obtain the VEV i.e. $\langle \phi^0 \rangle= v/\sqrt{2}$, 
the light neutrino mass matrix is given by}
{
\be
\label{neutrinomass}
(m^\nu)^{ab} = -\frac{v^2}{2}\left( \frac{y^a_T y^b_T}{m_T} + 
\frac{y^a_S y^b_S}{m_S}
\right)
\ed
The unitary PMNS matrix that diagonalizes the neutrino mass matrix Eq.~(\ref{neutrinomass}) is given by
\be
\label{pmns}
U=
\left(
\begin{array}{ccc}
 c_{12} c_{13} & s_{12} c_{13} & s_{13}e^{-\text{i$\delta $}} 
   \\
 - s_{12}c_{23} -c_{12} s_{23} s_{13} e^{\text{i$\delta $}}&
   c_{12} c_{23}-s_{12} s_{23} s_{13} e^{\text{i$\delta $}} &
    s_{23} c_{13} \\
 s_{12} s_{23}- c_{12} c_{23} s_{13} e^{\text{i$\delta $}}&
   -c_{12} s_{23} -s_{12} c_{23} s_{13} e^{\text{i$\delta $}}&
    c_{23} c_{13}
\end{array}
\right)\times \text{diag} (1,e^{i \Phi}, 1) .
\ed
where, $s_{ij} = \sin\theta_{ij}$, $c_{ij} = \cos\theta_{ij}$ (i,j = 1,2,3), $\delta$ is the CP-violating Dirac phase and $\Phi$ denotes the Majorana phase.  
The experimental constraints on the neutrino masses and mixing parameters,
at 2$\sigma$ level~\cite{Schwetz:2007my, Strumia:2006db} are
\be
7.3 \times 10^{-5} {\rm eV^2} < \Delta m^2_{S} &<& 8.1 \times 10^{-5} {\rm eV^2}\\
2.1 \times 10^{-3} {\rm eV^2} < |\Delta m^2_{A}| &<& 2.7 \times 10^{-3} {\rm eV^2}\\
0.28  < \sin^2\theta_{12} < 0.37, \qquad
0.38  < \sin^2\theta_{23} &<& 0.68, \qquad \sin^2\theta_{13} < 0.033\\\nonumber
\ed
In this section, we focus mainly on the case of Normal Hierarchy (NH, $m^\nu_1 = 0$),  and  Inverted Hierarchy (IH, $m^\nu_3 = 0$) neglecting the Majorana phase. Using the above experimental constraints, the neutrino masses
are given by~\cite{Schwetz:2007my, Strumia:2006db}
\be
 m^\nu_2 = \sqrt{\Delta m^2_{S}}, \qquad  m^\nu_3 = \sqrt{\Delta m^2_{S} + \Delta m^2_{A}}
\ed
in the case of NH, and 
\be
 m^\nu_1 = \sqrt{\Delta m^2_{A} - \Delta m^2_{S}}, \qquad  m^\nu_2 = \sqrt{\Delta m^2_{A}}
\ed
in the case of IH.
%
%
The constraints on the neutrino mass matrix elements direclty translate into the physical Yukawa couplings $y^a_{T}$. Using Casas-Ibarra parametrization~\cite{Ibarra:2003up}, one can find a formal solution for the Yukawa couplings $y^a_T$ and $y^a_S$ can be 
expressed as
\be
\label{yukawa1}
y^a_T &=& -i\frac{\sqrt{2\,m_T}}{v}
\bigg(\sqrt{m^{\nu}_2} \cos z\,U^{*}_{a2} + \sqrt{m^{\nu}_3} \sin z\,U^{*}_{a3}
\bigg)\\
y^a_S &=& -i\frac{\sqrt{2\,m_S}}{v}\bigg(-\sqrt{m^{\nu}_2} \sin z\,U^{*}_{a2} + \sqrt{m^{\nu}_3} \cos z\,U^{*}_{a3}\bigg) 
\ed
for NH, and
\be
y^a_T &=& -i\frac{\sqrt{2\,m_T}}{v}\bigg(\sqrt{m^{\nu}_1} \cos z\,U^{*}_{a1} + \sqrt{m^{\nu}_2} \sin z\,U^{*}_{a2}\bigg)\\
y^a_S &=& -i\frac{\sqrt{2\,m_S}}{v}\bigg(-\sqrt{m^{\nu}_1} \sin z\,U^{*}_{a1} + \sqrt{m^{\nu}_2} \cos z\,U^{*}_{a2}\bigg)
\label{yukawa2} 
\ed
for IH,
where $a=1,2 ,3$ and $z$ is a complex parameter. 
In order to study the effect of the $z$ parameter, 
 we show in Fig.1 the relative size 
 of the Yukawa couplings $y^a_T$ as a function 
of ${\rm Im}(z)$ for IH (left panel) and NH (right panel) 
with fixed $m_T=$ 1 TeV, when ignoring the influence of the
Majorana phase $\Phi$. 
As we can see from both panels, for large ${\rm Im}(z)$, 
the Yukawa couplings 
remain large and even ${\cal{O}}(1)$,
which can be understood from Eqs.~(\ref{yukawa1})-(\ref{yukawa2}) where $y^{a}_{T}$ are proportional to $e^{{\rm Im}(z)}$.
This allows to account for the experimental values of neutrino
masses without fine-tuning the Yukawa couplings
due to a cancellations in combinaition of them,
%
the observable effects are then possible.
%
%
}
\begin{figure}[h!]
\label{fig1:imz}
\vspace{1cm}
\begin{center}
\begin{picture}(280,220)
\put(-70,0){\mbox{\psfig{file=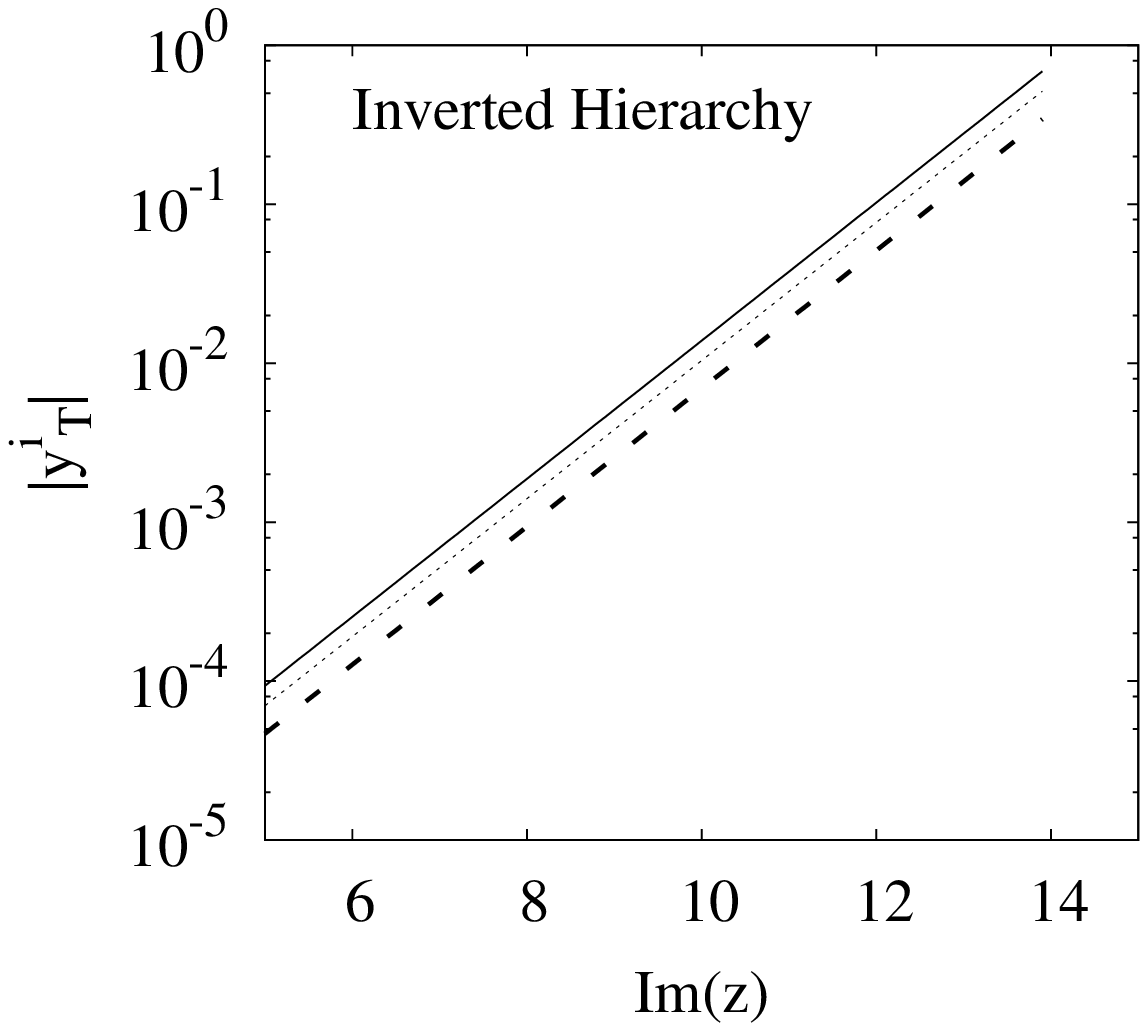,height=3.4in,width=3.0in}}}
\put(140,0){\mbox{\psfig{file=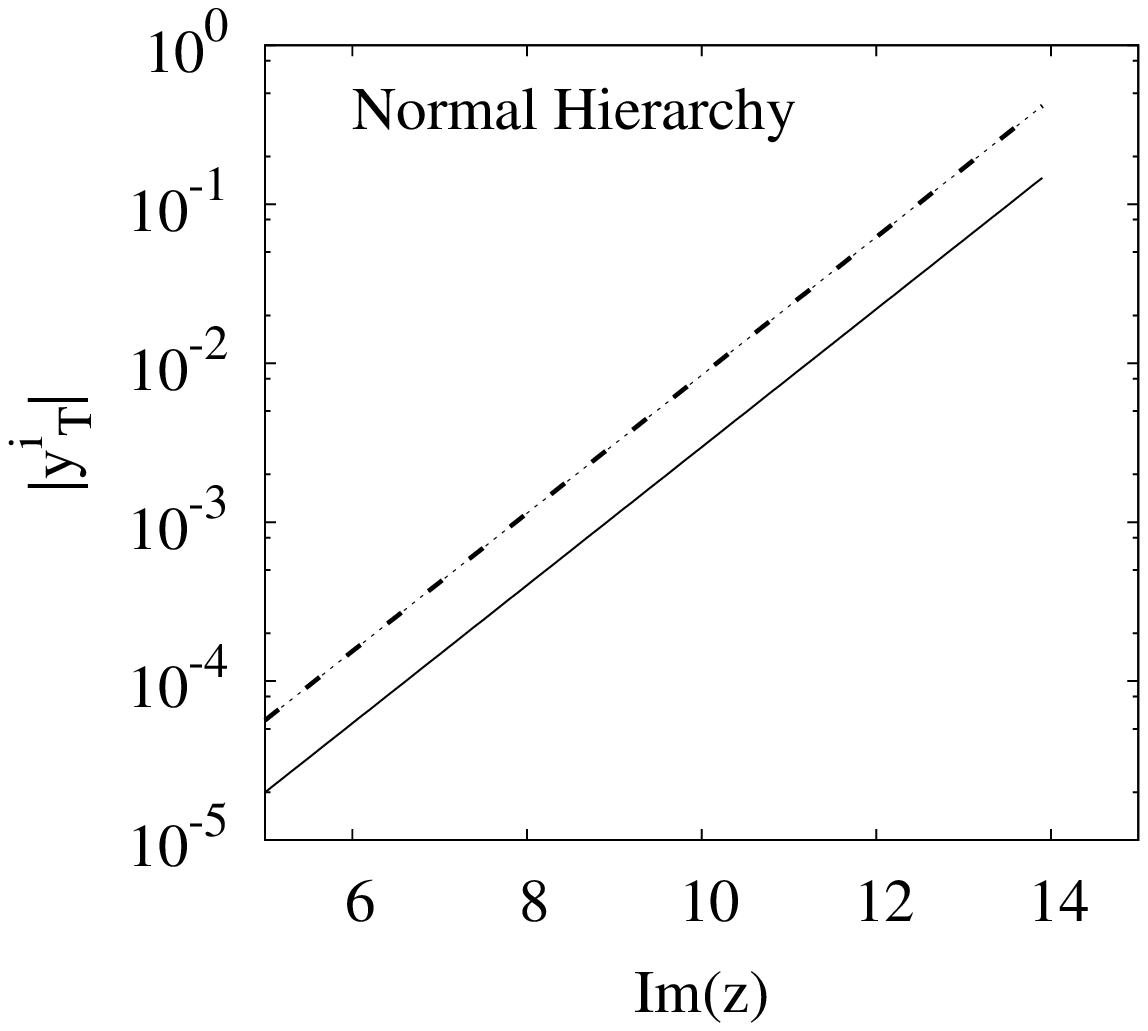,height=3.4in,width=3.0in}}}
\end{picture}
\end{center}
\caption{The absolute value of Yukawa couplings $y^a_T$ as a function of {\rm Im}(z) for IH (left) and NH (right), with $m_T = 1$ TeV and $\Phi =0$, in which the solid, dashed and dotted lines represent a=1,2, and 3,
respectively.}

\end{figure}

\section { Decay rates for $\tau\to \ell M$ and $\tau\to 3\ell$ }
\label{decay rates}
Based on previous analysis, now we can study the LFV of the type-III seesaw
model on semileptonic  $\tau\to \ell M$ with $M=(P, V)$ and leptonic
$\tau\to 3\ell$ decays. According to the interactions in
Eq.~(\ref{eq:Z}) and Z couplings in the SM, the relevant effective
Hamiltonian for $\tau$ flavor changing decays can be written as
 \be
 {\cal{H}} =
\frac{4 G_F}{\sqrt{2}} \cos2\theta_W Z_{Li3} \left( g^f_L\bar {f}
\ga^\mu P_L f \bar\ell_i \gamma_\mu P_L \tau + g^f_R \bar {f}
\ga^\mu P_R f \bar\ell_i \gamma_\mu P_L \tau \right)
\label{eq:4fermi}
 \ed
with
 \be
 g^f_L &=& I_{3f} - Q_{f} \sin^2\theta_W \,,\non\\
 g^f_R &=& -Q_{f} \sin^2\theta_W\,,
 \ed
where  we have used the equalities $\cos\theta_W = m_W/m_Z$ and
$g^2/8m^2_W = G_F/\sqrt{2}$, $f$ could be leptons and quarks, and
$I_{3f}$ and $Q_{f}$ denote the third component of weak isospin and
electric charge of the particle $f$, respectively. Since
semileptonic $\tau$ decays are associated with meson production
where the decay constants of nonperturbative hadronic effects
involve, for dealing with the hadronic effects, as usual the decay
constants of pseudo-scalar (P) and vector (V) mesons are defined as
 \be
\langle 0 | \bar q \ga^\mu \ga_5 q|P(p)\rangle &=& if_P
p^\mu\,,\non\\
\qquad \langle 0 | \bar q \ga^\mu  q|V(p)\rangle &=& i m_V f_V
\varepsilon^\mu_{V}
 \ed
with $\varepsilon^\mu_{V}$ being the polarization vector of vector
meson. Moreover, for the modes associated with $\eta$ and
$\eta^{\prime}$ mesons, we employ the quark-flavor scheme in which
$\eta$ and $\eta'$ physical states are described by
\cite{flavor0,flavor}
\begin{eqnarray}
\left( {\begin{array}{*{20}c}
   \eta   \\
   {\eta '}  \\
\end{array}} \right) = \left( {\begin{array}{*{20}c}
   {\cos \phi } & { - \sin \phi }  \\
   {\sin \phi } & {\cos \phi }  \\
\end{array}} \right)\left( {\begin{array}{*{20}c}
   {\eta _{q} }  \\
   {\eta _{s} }  \\
\end{array}} \right) \label{eq:flavor}
\end{eqnarray}
with $\phi$ being the mixing angle, $\eta _{q}  = ( {u\bar u + d\bar
d})/\sqrt{2}$ and $\eta_{s} = s\bar s$. Accordingly, the decay
constant of $\eta^{(\prime)}$ associated with $\bar q \ga^{\mu}
\ga^{5}q$ ($q=u,\ d$) current is given by
$f_{\eta^{(\prime)}}=\cos\phi( \sin\phi) f_{\eta_q}$.
Consequently,  for $\tau\to \ell_i P$ process, the decay amplitude
can be summarized by
 \be
\la P \ell_i| {\cal H}|\tau \ra &=& \sqrt{2}G_F f_P m_\tau
\cos2\theta_W Y_{P} Z_{Li3} \bar\ell_i P_R \tau\,,
 \ed
while for $\tau\to \ell_i V$ decay, it is
  \be
\la V \ell_i| {\cal H}|\tau \ra &=& \sqrt{2}G_F f_V m_V
\cos2\theta_W Y_{V} Z_{Li3} \bar\ell_i \not\varepsilon_V P_L \tau\,,
 \ed
where  $f_{\eta^{\prime}}=\sin\phi(\cos\phi) f_{\eta_s}$,
 \be
Y_{\pi^0} &=& -\frac{1}{\sqrt{2}}\,, \qquad  Y_\eta =-\frac{1}{2}
\,, \qquad Y_{\eta^\prime}
= \frac{1}{2} \,,\non \\
Y_{\rho^0} &=& \frac{\cos2\theta_W}{\sqrt{2}}, \quad Y_{\omega} =
-\frac{2}{3\sqrt{2}} \sin^2\theta_W , \quad Y_{\phi} =
-\frac{1}{2}+\frac{2}{3}\sin^2\theta_W \,.
 \ed
Due to $m_{e,\mu}\ll m_\tau$, we have neglected the masses of light
leptons. Hence, the BRs for semileptonic $\tau$ LFV are found to be
 \be
{\cal B}(\tau \to \ell_i P) &=& \frac{G^2_F}{16\pi\Gamma_\tau}
f^2_{P} m^3_{\tau}\cos^22\theta_W Y^2_P |Z_{Li3}|^2
 \left( 1 - \frac{m^2_P}{m^2_\tau}\right)^2\,,
\label{eq:B1}
 \ed
\be {\cal B}(\tau \to \ell_i V) &=&
\frac{G^2_F}{16\pi\Gamma_\tau}f^2_{V} m^3_{\tau}
\cos^22\theta_WY^2_V |Z_{Li3}|^2
 \left( 1 - \frac{m^2_V}{m^2_\tau}\right)^2 \left(
1 +2 \frac{m^2_V}{m^2_\tau}\right)\,. \label{eq:B2}
 \ed

For leptonic $\tau\to 3\ell$ decays, although there involve no
hadronic effects, however, they are three body decays and have more
complicated phase space. To simplify the formulation, we neglect the
effects of light lepton masses. Hence, using the interactions in
Eq.~(\ref{eq:Z}), the BR for $\tau\to 3 \ell$ is given by
 \be
 {\cal B}(\tau \to \ell_i \ell \bar\ell) =
 \cos^2 2\theta_W |Z_{Li3}|^2 \left( \zeta |g^\ell_L|^2 + |g^\ell_R|^2\right){\cal
B}(\tau \to \ell \nu_\tau \bar\nu_{\ell}) \label{eq:tau3l}
 \ed
where $\zeta = 2$ for $\ell_i = \ell$ and $\zeta = 1$ for $\ell_i
\neq \ell$.


After introducing the contributions of Z-mediated LFV in the
type-III seesaw model, in order to constrain the free parameters of
$Z_{Li3}$, we have to find out the possible strict limits. As known
that the Z-mediated effects at tree level in the SM are flavor
conserved, intuitively the flavor violating decays $Z\to \ell_i \bar
\ell_j$ with $i\neq j$, $\tau\to \ell (P, V)$ and $\tau\to 3\ell$
etc could give strong constraints on the unknown parameters.
Therefore, by examining the relation of the BR and the associated
parameter, one can easily find
$|Z_{Lij}|=|\Delta_{Li4}\Delta^*_{Lj4}|\propto \sqrt{BR}$. By taking
${\cal B}(Z\to \ell_i \bar\ell_j)\sim 10^{-6}$ and ${\cal B}(\tau\to
[\ell (P, V),\, 3\ell])\sim10^{-7}$, we see
$|\Delta_{Li4}\Delta^*_{Lj4}|\propto 10^{-3}$ and
$|\Delta_{Li4}\Delta^*_{Lj4}|\propto 10^{-4}$, respectively.
However, if we further think, the current high precision
measurements of $Z\to \ell_i \bar\ell_i$ processes in fact will
provide more severe limits. Roughly speaking, the key reason can be
understood by which the allowed range for new physics is directly
governed by small errors of data for $Z\to \ell_i \bar\ell_i$, {\it
i.e.} $|\Delta_{Li4}|^2\propto \Delta \Gamma/\Gamma_Z\equiv
[\Gamma^{\rm exp}(Z\to \ell_i \bar\ell_i) - \Gamma^{\rm SM}(Z\to
\ell_i \bar\ell_i)]/\Gamma_Z \sim 10^{-5}$ \cite{PDG08}. Although
the constraint in the real situation will depend on the detailed
characters of the process, however, we will adopt current data of
$Z\to \ell_i \bar\ell_i$ with $1\sigma$ errors as the inputs and the
BRs for $\tau\to \ell (P, V)$ and $\tau\to 3\ell$ decays are our
predictions.

In terms of the Z couplings in Eq.~(\ref{eq:Z}), the BR for $ Z \to
\ell_i \bar\ell_i $ decay with new effects  can be formulated as
\begin{eqnarray}
{\cal B} (Z \to \ell_i \bar\ell_i) &=& {\cal B}^{SM} (Z \to \ell_i
\bar\ell_i) + \xi_Z |\Delta_{Li4}|^2
\end{eqnarray}
with
\begin{eqnarray}
{\cal B}^{SM}(Z \to \ell_i \bar\ell_i ) &=& \frac{G_F m^3_Z}{3
\sqrt{2}\pi \Gamma_Z} \left[ \left(\frac{\cos2\theta_W}{2}\right)^2
+ \sin^4\theta_W\right]\,,\non\\
\xi_Z&=& \frac{m^3_Z G_F}{6 \sqrt{2}\pi\Gamma_Z} \cos^22\theta_W
\chi_L\,.
\end{eqnarray}
Due to $m_\ell\ll m_Z$, here we have dropped the mass of lepton. In
addition, we also neglected the terms that power in free parameter
is higher than $|\Delta_{Li4}|^2$. As a result, the allowed range
for unknown parameter can be bounded by
 \be
|\Delta_{Li4}|^2 &=& \frac{ \Delta{\cal B}_{i} }{\xi_Z }=\frac{1
}{\xi_Z }\left[{\cal B}^{\rm exp}(Z \to \ell_i \bar\ell_i ) - {\cal
B}^{\rm SM}(Z \to \ell_i \bar\ell_i )\right] \,.
 \ed
By using $G_F = 1.16634 \times 10^{-5}$ GeV$^{-2}$, $\sin^2\theta_W
= 0.223$ and $\Delta {\cal B}_{e,\, \mu,\, \tau}=(4,\, 7,\, 8)\times
10^{-5}$ where the values are taken from $1\sigma$ errors of data
for $Z\to \ell_i \bar\ell_i$ \cite{PDG08}, the upper limit on
$|\Delta_{Li4}|$ is given in Table~\ref{tab:lim}.
\begin{table}[hptb]
\caption{ Upper limit on $\Delta_{Li4}$ with $1\sigma$ errors of
${\cal B}(Z\to \ell_i \bar \ell_i)$. } \label{tab:lim}
\begin{ruledtabular}
\begin{tabular}{cccc}
Mode  & $ e^- e^+$ & $ \mu^- \mu^+$ & $\tau^-\tau^+$
\\ \hline
 $|\Delta_{Li4}|$ & $0.016$ & $0.021$ & $0.023$
\end{tabular}
\end{ruledtabular}
\end{table}
Thus, based on Eqs.~(\ref{eq:B1}) and (\ref{eq:B2}), the upper limits
on the BRs for $\tau\to \ell(\pi^0,\, \eta,\, \eta^\prime)$ and
$\tau\to \ell(\rho^0,\, \omega,\, \phi)$ are shown in Tables
\ref{tab:tau1} and \ref{tab:tau2}, respectively. Here, we have used
the hadronic values as
 \be
 && f_{\pi}=0.13,\ \ \ f_\eta=0.11, \ \ \ f_\eta'=0.135,\\ \non
 && f_\rho=0.216,\ \ \ f_\omega=0.187, \ \ \ f_\phi=0.237
 \ed
in unites of GeV. From the values in the tables, we see clearly that
the upper limits on the BRs for 
$\tau\to \ell (\pi^0,\, \rho^0,\, \phi)$ could be
${\cal{O}}(10^{-8})$ and the order in size is 
${\cal B}(\tau\to \ell \pi^0)>
{\cal B}(\tau \to \ell \rho^0)> {\cal B}(\tau\to\ell \phi^0)$.
Therefore, the semileptonic $\tau\to \ell (\pi^0,\, \rho^0)$ decays
could be the good candidates to probe the Z-mediated $\tau$ LFV.

\begin{table}[hptb]
\caption{ Upper limits on the BRs (in units of $10^{-8}$) for $\tau \to \ell (\pi^0,
\eta, \eta')$ decays with $1\sigma$ errors of ${\cal B}(Z\to \ell_i
\bar \ell_i)$ as the constraints. }\label{tab:tau1}
\begin{ruledtabular}
\begin{tabular}{cccc}
Mode & $\tau\to (e,\, \mu) \pi^0$ & $\tau\to (e,\, \mu) \eta$ &
$\tau\to (e,\, \mu) \eta^{\prime}$
 \\ \hline
Current limit & $(8.0,\, 11)$ & $ (9.2,\, 6.5)$ & $(16,\, 13)$ \\
This work& $(2.8,\, 5.0)$ & $(0.6,\, 1.0)$ & $(0.4,\,0.7)$
\end{tabular}
\end{ruledtabular}
\end{table}
\begin{table}[hptb]
\caption{ Upper limits on the BRs (in units of $10^{-8}$) for $\tau \to \ell (\rho^0,
\omega, \phi)$ decays with $1\sigma$ errors of ${\cal B}(Z\to \ell_i
\bar \ell_i)$ as the constraints. }\label{tab:tau2}
\begin{ruledtabular}
\begin{tabular}{cccc}
Mode & $\tau\to (e,\, \mu) \rho^0$ & $\tau\to (e,\, \mu) \omega$ &
$\tau\to (e,\, \mu) \phi$
 \\ \hline
Current limit & $(6.3,\, 6.8)$ & $(11,\,8.9)$ &
$(7.3,\, 13)$ \\
This work & $(1.7,\,3.0)$ & $(0.09,\,0.2)$ & $(1.1,\,1.7)$
\end{tabular}
\end{ruledtabular}
\end{table}

With the same constraints shown in the Table~\ref{tab:lim}, the
values of BRs for leptonic $\tau$ decays formulated by
Eq.~(\ref{eq:tau3l}) are presented in Table \ref{tab:tau3}. It is
clear that the BRs for all $\tau\to 3\ell$ decays are of order
$10^{-8}$ and the predictions are close to each other. Furthermore,
from the Table \ref{tab:tau3}, one can find that the value of ${\cal B}(\tau\to
3\mu)$ is a little bit larger than current experimental upper limit.
It seems that $\tau\to 3\mu$ provides the most strictest constraint
on the free parameters. However, by reexamining the constraints of
$Z\to \ell_i \bar\ell_i$, we find that the reverse situation is
arisen from the errors of $Z\to (\tau^-\tau^+, \mu^- \mu^+)$ being
larger than that of $Z\to e^- e^+$, i.e. $\Delta {\cal B}_{\tau}\sim
\Delta {\cal B}_{\mu} > \Delta {\cal B}_{e}$. If we adopt $3\sigma$
of the world average ${\cal B}(Z\to \ell_i \bar\ell_i)=(3.3658\pm
0.0023)\%$ for $\ell=e,\, \mu$ and $\tau$,
{the new upper limits on the BRs for semileptonic and
leptonic decays are found to be}
 \be
 && {\cal B}[\tau\to \ell (\pi^0,\, \eta,\, \eta')]<(4.2,\, 0.8,\,
0.6 )\times 10^{-8}\,,\non \\
 && {\cal B}[\tau\to \ell (\rho^0,\,
\omega,\, \phi)]<( 2.5,\, 0.1,\, 1.6)\times 10^{-8}\,, \non\\
 && {\cal B}[\tau\to  (3\ell, \mu e^- e^+,\, e\mu^-\mu^+)]<( 3.1,\, 2.0,\, 2.0)\times
 10^{-8}\,.
 \ed
Clearly, precision measurements of $Z\to \ell_i \bar\ell_i$ play an
essential role on the constraints.

\begin{table}[hptb]
\caption{ Upper limits on the BRs (in units of $10^{-8}$) for $\tau \to 3\ell$ decays
with $1\sigma$ errors of ${\cal B}(Z\to \ell_i \bar \ell_i)$ as the
constraints.}\label{tab:tau3}
\begin{ruledtabular}
\begin{tabular}{ccccc}
Mode & $\tau\to 3e$& $\tau\to 3\mu$ & $\tau\to \mu e^- e^+$ &
$\tau\to e \mu^- \mu^+ $
 \\ \hline
Current limit & $3.6$ &$3.2$ & $2.7$ & $3.7$\\
This work &  $2.1$ & $3.7$ & $2.3 $ & $1.3 $
\end{tabular}
\end{ruledtabular}
\end{table}


\section {Conclusions}\label{conclusions}

We have investigated the lepton flavor violating effects in the
framework of type-III seesaw model by extending the SM with one $SU(2)_L$
triplet and singlet fermions. 
Due to the difference in weak charges between new and ordinary leptons, 
intriguingly Z-mediated LFV is
generated at tree level. 
Moreover, it is found that the significant
effects only occur in the left-handed leptons. Although LFV could be
induced by charged currents through one-loop, however, comparing
with tree contributions, they are subleading effects and neglected
in our analysis. To illustrate the novel effects, we study the
semileptonic $\tau\to \ell M$ and leptonic $\tau\to 3\ell$ decays.
For numerical calculations, we find that the precision measurements
of $Z\to \ell_i \bar\ell_i$ play an important role on the
constraints of the free parameters. 
Furthermore, we find that the upper limits on
the BRs for $\tau\to \ell (\pi^0,\, \rho^0,\, \phi)$ and $\tau\to
3\ell$ could reach ${\cal{O}}(10^{-8})$ in the model under discussion.
%





\section{acknowledgments}
We would like to thank Carla Biggio for useful discussions and Goran Senjanovi\'c for illuminating discussions during the "Signaling the arrival of the LHC era" workshop in ICTP where this work was initiated.
R.B is supported by National Cheng Kung University Grant No. HUA
97-03-02-063 and
C.C.H is supported by the National Science Council of
R.O.C under Grant \#s: NSC-97-2112-M-006-001-MY3.


\end{document}